\documentclass[letterpaper]{article}
\usepackage[dvips]{graphicx}
\begin{document}

\title{The effect of social interactions in the primary consumption life cycle of motion pictures}
\author{C\'esar A. Hidalgo R.$^{12}$, Alejandra Castro$^{3}$, Carlos Rodriguez-Sickert$^{45}$}
\date{}
\maketitle

\begin{center}
\textit{$^1$ Department of Physics and Center for Complex
Network Research, University of Notre Dame, Notre Dame, In 46556.\\
$^2$ Kellogg Institute, 130 Hesburgh Center, Notre Dame, In, 46556.\\
$^{3}$ Department of Physics, University of Michigan, Ann Arbor
48104.\\
$^{4}$ Department of Sociology, Pontificia Universidad Cat\'olica de
Chile, Vicu\~na Mackenna 4860, Macul, Santiago Chile.\\
$^{5}$Santa Fe Institute, 1399 Hyde Park Road, Santa Fe, NM 87501.
USA
 }

\end{center}

\begin{abstract}
We model the consumption life cycle of theater attendance for single
movies by taking into account the size of the targeted group and the
effect of social interactions. We provide an analytical solution of
such model, which we contrast with empirical data from the film
industry obtaining good agreement with the diverse types of
behaviors empirically found. The model grants a quantitative measure
of the valorization of this cultural good based on the relative
values of the coupling between agents who have watched the movie and
those who have not. This represents a measurement of the observed
quality of the good that is extracted solely from its dynamics,
independently of critics reviews.
\end{abstract}

\section{Introduction}

The study of the consumption of cultural goods in general, and that
of films in particular, has been traditionally restricted to total
demand empirical studies. Most of these studies have followed the
original guidelines set by the earliest authors, such as Baumol \&
Bowen \cite{baumol} or Moore \cite{moore}. In these studies
variations in quality of the good have been consigned to a residual
status, focusing in the effect of prices and income as the main
explanatory factors. Other studies, such as the one performed by
Blanco \& Ba\~nos-Pino \cite{blanco}, have considered the
availability of alternative leisure activities; while on the other
side, models such as the one presented by Thorsby \cite{thorsby}
have considered the impact of quality on final attendance,
incorporating it either as an expected value at the individual level
or as a macro variable obtained ex-post from critic's review indexes
\cite{urrutiager} or from online reviews \cite{chevalier}. Although
we acknowledge their efforts, we must say that in the latter works
the authors were not concerned with the structure of the consumption
cycle as an indicator of film quality. Their work is close to ours,
only in the sense that reviews act as a form of social interaction,
which turns out to be a key ingredient of our model.

However useful, demand studies as the ones surveyed above do not
account for the dynamics of the consumption life cycle of the
cultural good and, more importantly, they do not discuss the ways in
which the structure of this cycle is affected by the transmission of
information from members who have had access to the cultural good to
potential consumers.

Beside economic analysis, the sociologist Lipovetsky
\cite{lipovetsky} has focused his work on the macro structure of the
life cycle, mainly in terms of its duration. Basic principles which
underlie this process, however, are not identified in his work. In
this paper we propose a model based on first principles, which
agrees with all the observed behaviors exhibited by the empirical
data of the movie industry. The model is concerned with the primary
life cycle of the good, but it differs from the others in the sense
that the whole structure of the life cycle is recovered, not only
the characteristic decay times or the total consumption are
considered like in previous studies.

Although the consumption life cycle of some cultural goods can be
potentially unbounded (we still buy copies from Don Quixote), The
aim of this work is to understand the dynamics of the primary life
cycle, which finishes when per period consumption decreases below a
certain threshold relative to its premiere level. A less ambiguous
case is that of performing arts like theater, where producers are
forced to cancel presentations when box-office revenues goes below
fixed costs. The alternative cost associated with the availability
of new options, effect which also applies to the film industry or
best-sellers book industry, strengthens the limited duration life
cycle that characterizes aggregated consumption in these
environments. We present a model that reproduces the life cycle
dynamics and is determined by three basic factors:

\begin{itemize}
\item (i) the size of the targeted group
\item (ii) the prior conception about the quality of the good in question; and
\item (iii) the effect of social interactions, in the form of
information about the quality of the good, between agents who have
effectively experienced the good and potential consumers.
\end{itemize}

It will be the latter effect the one which determines the particular
shape of the consumption life cycle.

In relation with the literature studying the diffusion of
technological innovations, our model relates to the works of
Mansfield \cite{mansfield1,mansfield2}, and Bass \cite{bass}, with
the difference that the dynamical process that underlies the shape
of the cycle are identified and made accountable for it. Whereas
Bass uses a parameterized curve that he fits into the empirical
behavior, without incorporating any dynamical justifications for the
curves he chooses. Word of mouth can be justified as a valid
mechanism of social influence by following works like the one
presented by Moore \cite{moore2}. He argues that given the scale
economy that characterizes certain technologies, e.g., database
software, it is impractical (or too expensive) to do small-scale
experiments and, thus, word-of-mouth becomes an important part of
the diffusion process of these investment goods.

Going to the movies, however, poses a somewhat different problem for
potential consumers. First, "early adopters" do not face the risks
normally associated with the adoption of a novel technology and thus
an important fraction of them will be willing to consume the good
before peer's opinions become available. This will be performed
based solely on the pre-opening expectations generated by the media
and the producers themselves. This leads the life-cycle consumption
process to diverge from the standard S-shaped behavior associated
with the adoption of a novel technology. On the other hand, new
movies which become available for the general public tend to have an
initial attendance which is relatively high and usually decrease all
along the consumption cycle, the sole exception being small
productions or independent films which show a consumption life-cycle
that resembles that of novel technology adoption. Our model captures
both forms of behavior.

Other family of models, which takes into account social interaction
effects in the dynamics of consumption, is that of "fashion cycles"
\cite{karni,pesendorfer}. In our case, however, the demand of a good
does not decrease due to the consumption of it by other agents, the
movie does not become worn out, like a fashion design, in fact what
the model attempts to capture is the fact that agents who have seen
the movie have an impact on the expected value of the good of agents
who have not seen it.

A different approach for the diffusion of innovations is to consider
site percolation on regular lattices \cite{stauffer1,goldenberg}.
These models are based on the assumption that agents occupy the
vertex of a regular lattice in $d$ dimensions and are represented by
a random number. Innovations diffuse by percolating the lattice
according to a certain quality value which when above the
percolation threshold generates a giant consumption cluster.
Although we acknowledge this efforts we believe that spatially
realistic models should consider a network substrate instead of a
regular lattice. Social networks exhibit topological properties
which are completely different to the substrate defined by a regular
lattice, such as the scaling properties of its degree distribution
\cite{laszlo1,laszlo2}, the community structure
\cite{newman1,newman2} and the short average path length
\cite{watts} to name a few. In this paper we present a model that
does not take this substrate into account, it could be consider a
mean field solution or jelly model, regardless of this we are able
to reproduce all the types of aggregated behavior.


\section{The Model}

We now present a model of cultural consumption based on the
following assumptions.

\begin{itemize}
\item (i) Each agent goes to see a movie at the theater only once.
(We neglect the probability of going twice assuming that the
probability of going to the theater to see the same movie  $n$ times
is a rapidly decaying function of  $n$).
\item (ii) The probability
that an agent goes to the movies is affected by the interactions
with agents who have already seen it.
\end{itemize}

The different quantities associated with the model will be expressed
using the following notation.  $N(t)$   will represent the number of
agents that have not watched the movie at discrete time $t$, while
$P(t)$ will represent the probability that an agent who has not
watched the movie decides to attend it at discrete time $t$. The
observable quantity that can be measured is the number of agents who
have seen the movie at discrete time $t$. We will call this quantity
$A(t)$. In our model, it will be given by the product
$A(t)=P(t)N(t)$ , which is nothing more than the expected number of
agents attending the good.

Before the movie is available for its potential consumers, agents
have a prior conception about its quality, which comes from the
information of pre-observable features such as its budget, the cast,
and advertisement. The availability of this information will depend
on the marketing strategies of the producers and distributors.
Irrespectively of the way in which these strategies are
conceptualized, i.e., whether publicity is taken as an information
provider or as a persuasive device, its effects on our model are
equivalent, affecting the prior conception about the movie in
question and, therefore, the agents likelihood of actually
purchasing it at the beginning of the process. We will denote this
likelihood by $P_0$ , and we will call the initial target population
$N_{0}$ ; which represents the total number of potential attendants
of the good before it becomes available.

It can be argued that the length of the consumption cycle that the
paper takes into consideration is unrealistically short, since the
energy of film producers is directed to finding, successfully, new
ways of lengthening it. Sedgwick \cite{sedgwick} claims that more
than 70 per cent of film revenue is derived from non-theatrical
resources. However, given the dramatic effect that the primary cycle
of consumption has in the subsequent ones, we believe that this
primary cycle is worth of investigation on its own.

\subsection{An Atomized Society with no Supply Restrictions}

The simplest case is the one of an atomized society where agents'
decisions are independent from each other, both in terms of
information  the opinion of agents who have watched the movie do not
influence the decision of potential consumers  and in terms of
consumption  there are no network externalities associated with
timely coordinated consumption in this case, we will also assume
that restrictions on the supply side do not apply. The whole
targeted population could simultaneously go to the theater and the
horizon of its exhibition has no time limit.

In this case the probability of attending the theater does not
depend on time, and it is equal to its initial value $P_0$. The
system dynamics can be described by considering that the expected
attendance at a given time is given by the product between $P_0$ and
$N(t)$; and that at the beginning of the process no agents have seen
the movie. After the first time step some agents will have attended
the theater, and we will therefore have to subtract them from the
ones who have not. Thus the temporal variation of     will be given
by
\begin{equation}
N(t+1)=N(t)-P_{0}N(t) \label{1}
\end{equation}

If we approximate the discrete variables by continuous ones and
notice that in the first time step attendance is given by    , we
can conclude that the general solution of equation (1) has the
traditional form
\begin{equation}
N(t)=N_{0}e^{-P_{0}t}
\end{equation}

\begin{equation}
\label{3} A(t)=N_{0}P_{0}e^{-P_{0}t}
\end{equation}

in the next section we will see how social interactions alter this
behavior.

\subsection{Cultural Consumption and the Effect of Social
Interactions}

Once the effect of social interactions are considered,  $P(t)$
becomes a dynamical variable and its change in time is then due to
two main contributions. The first one is the one associated with the
transmission of information about the observed value of the movie,
and it represents the change in the probability of attending the
theater induced by the information about it transmitted between
agents. After the first period of attendance, residual potential
consumers have access to the opinion of the first period audience.
This effect is cumulative as the residual consumers have potential
access to the opinions of audiences from the previous time steps
\footnote{ In our model, the word-of-mouth process is one of first
order (only opinions of effective attendants are payed attention
to). Alternatively, one could consider one of  $n$   order where
opinions are transmitted to residual consumers also via agents who
have not yet participated in the consumption of the cultural good.
Qualitative properties of an  $n$   order process in which the
effect's strength diminishes over time, would be equivalent to the
$1^{st}$ order process investigated here. Wu et al.\cite{wu} have
shown that information flow remains bounded as 'long as the
likelihood of transmitting information remains sufficiently small} .
We assume that the change induced in $P(t)$ by this effect will be
proportional to the probability that agents who have not seen the
movie meet with agents who have. The proportionality constant
associated with this term will be the one representing the strength
of it. The second contribution we considered is the one associated
with coordinated consumption. In the case of performing arts, agents
usually attend in groups; therefore agents who have not seen the
movie are less likely to go if they are not able to find other
agents to keep them company. The change induced by this effect will
always reduce attendance likelihood, and it is also proportional to
the probability that agents who have attended the good meet with
ones who have not\footnote{The latter contribution to change can
also be associated with other social behaviors. An example of this
is that agents do not consume coordinately just because of company,
but can do so independently but at similar times, so they can
discuss the good in question latter.}. Therefore, when we consider
the effect of social interactions, we can write the change in the
probability of attending the good as

\begin{equation}
\label{4}
P(t+1)=P(t)+\sum_{t}(S_{+}(t)-S_{-}(t))\frac{A(t)}{N_0}\frac{N(t)}{N_{0}},
\end{equation}

where  $S_{+}$  represents the proportionality constant associated
with the effects of information flow while  $S_{-}$   is the one
associated with the social coordination effects. From now on we will
focus on the case in which $S_{+}$ can take positive and negative
values representing the fact that information transmitted between
agents can stimulate or inhibit the future attendance of other
agents. Whereas $S_{-}$ will always contribute negatively. This is
because coordinated consumption can only reduce the likelihood of
going to the theater for un-coordinated agents.  Both terms, it is
worth noticing, act directly on the population, altering the
likelihood that agents will attend or purchase the good. Although
agents are not explicit rational optimizers who make their decision
over the basis of a Bayesian update\footnote{ Ellison and Fudenberg
\cite{ellison} develop a Bayesian model of learning in which agents
update their beliefs about the quality of the product after meeting
other agents. In their model, it is the updated belief that affects
the agents' optimization process.}  of their beliefs about the
movie's quality, one could understand this model as the outcome of
rational searching behavior in an incomplete information
environment.

\subsubsection{Analytical Solution}

The continuous version of the system can be solved analytically in
the case in which  $S_{+}$,$S_{-}$   do not depend on time and the
effects are not cumulative\footnote{We make these assumption in
these part for technical reasons: It makes the problem analytically
tractable. It also satisfies the KISS principle (keep it simple):
although cumulative effects could be incorporated we observed that
there effect is to accelerate the process, which is equivalent to
time re-scaling. We could also argue that agents are more likely to
express there opinion soon after attending the picture, although our
main reason is that cumulative effects are not needed to make our
point, and therefore should not be incorporated if unnecessary.}. In
this case (1) and (4) reduce to
\begin{equation}
\label{5}
N(t+1)=N(t)-P(t)N(t)
\end{equation}
\begin{equation}
\label{6}
P(t+1)=P(t)+(S_{+}-S_{-})\frac{A(t)}{N_0}\frac{N(t)}{N_{0}},
\end{equation}
which in the continuous case can be represented by
\begin{equation}
\label{7}
\frac{dN}{dt}=-PN
\end{equation}
\begin{equation}
\label{8} \frac{dP}{dt}=\sigma P N^2
\end{equation}
where
$$
\sigma=\frac{S_{+}-S_{-}}{N_{0}^2}
$$

To find a solution we notice that equation (7) can be substituted
into equation (8) giving us a third differential equation relating
$N$ and $P$ which can be solved and used to introduce the initial
conditions of the system. This relation can be expressed as
\begin{equation}
P(t)=-\frac{\sigma}{2}(N(t)^2-N_{0}^2)+P_0. \label{9}
\end{equation}

Replacing \ref{9} back on equation (7) and integrating it, we find
that the solution of the system is given by
\begin{equation}
N^2(t)=\frac{BN_{0}^{2}}{N_{0}^2 - (N_{0}^2 - B)e^{B\sigma t}},
\end{equation}
\begin{equation}
P(t)=-\frac{\sigma}{2}\bigg( \frac{BN_{0}^2}{N_{0}^2 -
(N_{0}^2-B)e^{B\sigma t}} - N_{0}^2\bigg)+P_0
\end{equation}
where the constant  $B$ is defined as
$$
B=N_{0}^2+\frac{2P_0}{\sigma}
$$
\begin{figure}[t]
\begin{center}
\includegraphics[width=1\textwidth]{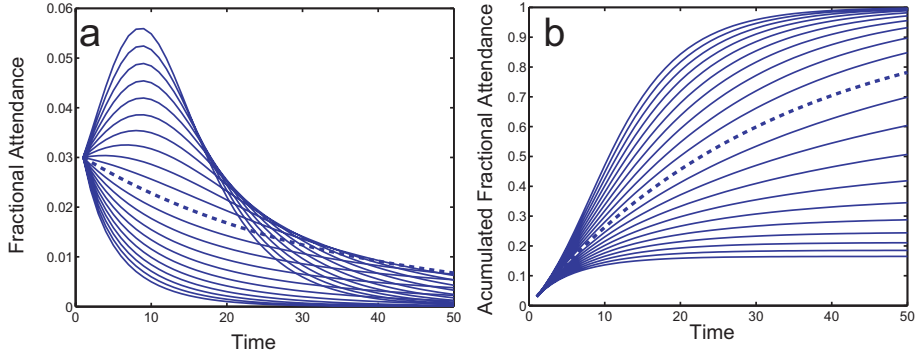}
\caption{\label{fig1} (a) Fraction of agents attending the theater a
a function of time according to the model described by eqns.
(\ref{7}) and (\ref{8}). The segmented line represents the atomized
behavior described by eqn. (\ref{3}) while the lines on top of it
and below of it represent the behavior of positive and negative
$\sigma$ values respectively. (b) Shows the accumulated attendance,
or in other words, the integral in time of (a).}
\end{center}
\end{figure}

Figure \ref{fig1} shows the attendance as a function of time as well
as the accumulated attendance normalized by the total target
population ( $A(t)/N_0$ and $\sum_{t}A(t)/N_0 $). The segmented line
represents the atomized behavior occurring when  $P(t)$   remains
constant throughout the process, it also divides the behavior of the
system in two, the solutions that lie above it are examples of cases
in which $\sigma$ is a positive quantity, these are examples of
systems which are dominated by a strongly favorable flow of
information stimulating the consumption of the good. On the
contrary, the lines that lie below the dashed one are the ones for
which $\sigma$ is a negative quantity and are dominated by
coordination effects. Summarizing, we can see that two classes of
behavior are predicted. The first class is characterized by a
monotonic decay with an exponential tail. Whereas, when the strength
of social interactions $\sigma$ is large enough, a second class of
behavior emerges. In this case a period of increasing attendance
exists until a maximum is reached. After this, the first class of
behavior develops. The latter case has an accumulated behavior that
resembles the standard ogive S-shape, which characterizes
technological diffusion.


\section{Empirical Validation}

The model represented by equations  (5) and (6) was validated by
comparing it with the U.S. box-office data available on the Internet
Movie Database (IMDb) Web site. We considered the 44 movies with the
highest budgets of 2003 as our sample set and performed a $\chi^2$
estimation procedure on all movies in order to find the parameter
set which most accurately fitted the empirical results.

The model has 3 free parameters, the two initial conditions $N_{0}$
and $P_{0}$, and the constant representing the strength of social
interactions $\sigma$. The parameters used in the $\chi^2$
minimization process were $\sigma$ and $N_0$ while $P_0$ was
determined by matching the first data point on the data set with the
first data point in the model-$P_0=(First\: Data \: Point)/N_0$-.
This reduced the number of free parameters to just two. The data set
considered did not contain weekly attendance as a function of time
but the weekly gross collected by the movie. We consider that the
amount of money collected in the box office by a movie is a linear
function of the number of people that attended it. This allow us to
do our empirical analysis based on the weekly amount of money
collected by a particular movie, which is equivalent to making the
analysis based on the number of agents attending it.
\begin{figure}[t]
\begin{center}
\includegraphics[width=1\textwidth]{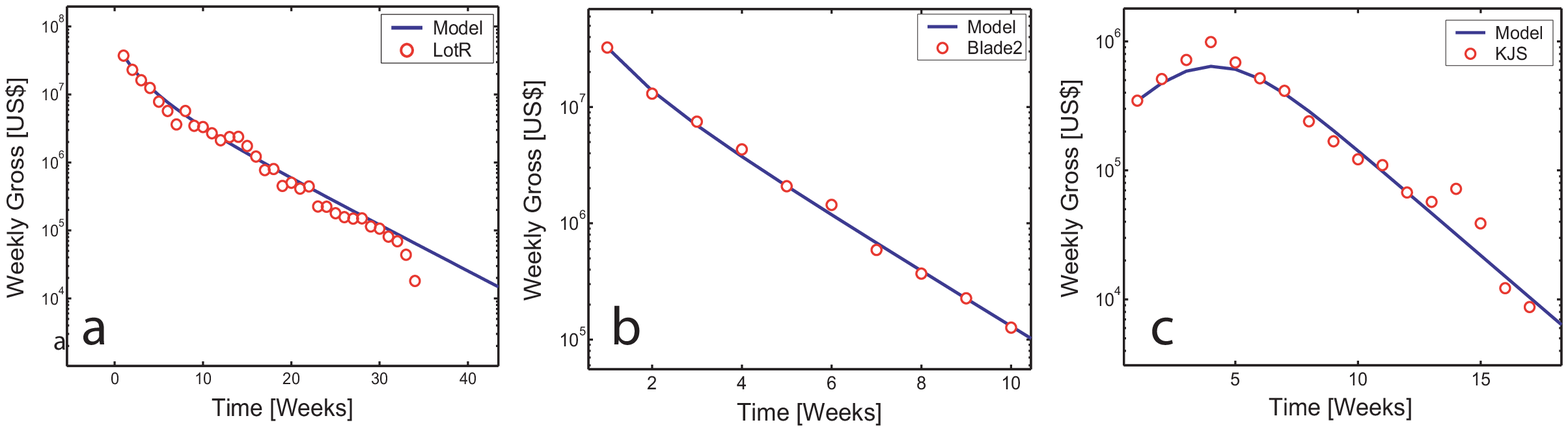}
\caption{\label{fig2} Three examples of the fitting procedure are
shown for (a) Lord of the Rings I (b) Blade II (c) Kissing Jessica
Stein.}
\end{center}
\end{figure}
\indent Figure (\ref{fig2}) shows three plots comparing the model
with the empirical data. Both, The Lord of the Rings: Fellowship of
the Ring, and Blade II represent two examples of the first class of
behavior identified in the previous section. It is worth noticing
that the slope followed by the data changes in time. The first
points have a negative slope, which is considerably more pronounced
than the ones in the tail. A simple exponential fit will be accurate
in only one section of the curve, but would fail to fit both
behaviors that were expected from the model and that appear to be
present in nature

In order to interpret the relative values of $\sigma$, and therefore
compare different motion pictures, it is important to consider the
behavioral class in which the particular films are
located\footnote{Other definition of behavioral class based on the
length of the consumption life cycle is given by \cite{sinha}.}.
Blockbuster movies have a high initial attendance, which causes
population finite size and social coordination effects to be very
strong. This tell us that movies in this category will usually have
a negative value for $\sigma$, and the magnitude of it will
represent the observed value of the good. Small negative values are
associated with movies in which the decay is not accelerated due to
the observed quality, but exhibit some acceleration due to social
coordination effects. On the other hand, large negative values map
into accelerated decays that we believe are due to social
coordination effects plus poor values in the observed quality of the
good.

In the second class of behavior, the premier level of attendance is
low, thus coordination effects do not act strongly on the system,
because of the initially slow depleting of potential customers. In
this case, small values of $\sigma$  represent a movie in which
agents attended in a random fashion and the decay is not accelerated
or damped due to social effects. On the other hand when relatively
large values of $\sigma$ are observed the attendance increases
during the first time steps. This increase in consumption is due to
the fact that the movie was well evaluated and that the pool of
possible attendants was not depleted during the first couple of
weeks.

Summarizing, in order to interpret the parameter associated with the
observed quality of the good $\sigma$, it is important to consider
two things. First we consider the class of behavior to which the
movie actually pertains, this can be done by observing if the
premiere attendance represents a large fraction of the total one.
Then, once the movie has been associated with a certain class of
behavior, the actual value of sigma should be interpreted relative
to the distribution of values associated with movies in that
particular class

\begin{figure}[t]
\begin{center}
\includegraphics[width=0.5\textwidth]{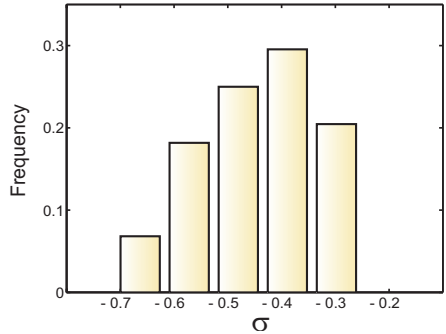}
\caption{Distribution of the social influence parameter $\sigma$ for
the 44 studied movies. \label{fig4}}
\end{center}
\end{figure}

From the empirical point of view and after observing the whole data
set it becomes apparent that the values of sigma that we found tend
to lie in a well defined distribution. Figure (\ref{fig4}) shows the
distribution of values for $\sigma$ obtained using this data set and
the procedure explained above. Given the selection criterion used,
movies with the highest budgets, it is natural to see that this
curve shows $\sigma$  values belonging to the blockbuster class of
behavior. A negative bias is observed when we look at the
distribution of $\sigma$ presented in figure (\ref{fig4}). This
negative bias tells us that in most cases, the coordination effect,
which could also be associated with the underlying effect of
cultural competition, dominates the dynamical properties of the
system.
\indent On the basis of the conjecture that coordinated consumption
has the same effect on all motion pictures inside the same
behavioral class, we can interpret that the value extracted for
$\sigma$ represents the actual effect that the flow of the
information associated with the observed value has on the film, and
therefore on its consumption life cycle. This would indicate that
coordinated consumption only introduces a shift in the value of
$\sigma$ and that the deviation from this well-defined mean
represents the actual value of the movie as given by the targeted
audience. For instance, the differences in the estimated parameter
for The Lord of the Rings and Blade II, would indicate that the
former film was much better received than the latter one.
\textbf{What is interesting about this result is that we do not
claim an actual value based on the opinion of a film critic or an
audience survey, we infer it from the structure of attendance
behavior we observe}. This reversed engineering, definition of the
observed value of subjective \emph{film quality}  presented here can
be used to characterize consumers' response to specific genres and
thus help target film publicity and distribution in a more effective
way.
\section{Conclusion}

A dynamical model representing the life cycle of motion pictures was
introduced. The assumptions of the model were that

\begin{itemize}
\item i) agents do not go to the theater to see a particular movie
more than once, and that \item ii) the probability that agents go to
the theater changes in a way which is proportional to the number of
agents who have attended the theater times the ones who have not.
\end{itemize}

The first of these assumptions is the one that gives rise to the
exponential decay that characterize the tail of this process, while
the second one is the one that allows the system to adjust its decay
according to the social interactions present in cultural
consumption, and consequently allows the model to accurately fit the
different classes of observed behavior, namely,

\begin{itemize}
\item i) a monotonic decay with an exponential tail, and
\item ii) an exponential  adoption followed by an exponential tail
which is traduced into an ogive S-shaped behavior when the
accumulated theater attendance is observed.
\end{itemize}

In addition, under certain assumptions, our model can be used to
infer a quantitative estimator of the subjective reception of a
particular film. An estimate which is independent of critics review
and is inferred only from the shape of the consumption life cycle.

Further research in this area can be directed in a variety of ways.
A natural extension of the model analyzed here would be to consider
the more general setting in which the agents face a number of
cultural options and a limited budget for a given period of time. It
is also an area interest to investigate the link between the primary
life of consumption and subsequent phases of consumption. Another
route for more empirically oriented research could be associated
with the investigation of longitudinal processes in order to explore
how the structure of the consumption life cycle has evolved along
the last decades, and transversal processes to see the differences
and correlations between the reception of particular films in
different geographical regions. A more complete analysis of the
movie industry database should be carried out in order to understand
the nature of these structures.

We thank the comments of Samuel Bowles, Peter Richerson, Marcos
Singer and the participants of the workshop on Evolution and the
Social Sciences at the CEU, Budapest. We also acknowledge financial
support from Fundacion Andes grant C-13960.

\end{document}